\documentclass[prd,aps,twocolumn,amsmath,amssymb,nofootinbib,preprintnumbers]
{revtex4}

\usepackage{graphicx}
\usepackage{dcolumn}
\usepackage{bm}
\usepackage{amsmath}
\usepackage{amsfonts}
\usepackage{euscript,bbm}
\usepackage{ifthen}
\usepackage{psfrag}
\usepackage{slashed}

\def\ls{\mathrel{\lower4pt\vbox{\lineskip=0pt\baselineskip=0pt
           \hbox{$<$}\hbox{$\sim$}}}}
\def\gs{\mathrel{\lower4pt\vbox{\lineskip=0pt\baselineskip=0pt
           \hbox{$>$}\hbox{$\sim$}}}}
\def\drawbox#1#2{\hrule height#2pt

\hbox{\vrule width#2pt height#1pt \kern#1pt
              \vrule width#2pt}
              \hrule height#2pt}

\def\Asym#1#2{\vcenter{\vbox{\drawbox{#1}{#2}
              \kern-#2pt       
              \drawbox{#1}{#2}}}}

\newcommand{\Expect}[1]{\left\langle #1 \right\rangle}
\newcommand{\be}{\begin{equation}}
\newcommand{\ee}{\end{equation}}
\newcommand{\bea}{\begin{eqnarray}}
\newcommand{\eea}{\end{eqnarray}}

\begin{document}

%
\title{Cladogenesis: Baryon-Dark Matter Coincidence from Branchings in Moduli Decay}

\author{Rouzbeh Allahverdi$^{1}$}
\author{Bhaskar Dutta$^{2}$}
\author{Kuver Sinha$^{2}$}

\affiliation{$^{1}$~Department of Physics and Astronomy, University of New Mexico, Albuquerque, NM 87131, USA \\
$^{2}$~Department of Physics, Texas A\&M University, College Station, TX 77843-4242, USA}

\begin{abstract}

We propose late-time moduli decay as the common origin of baryons and dark matter. The baryon asymmetry is produced from the decay of new TeV scale particles, while dark matter is created from the (chain) decay of $R$-parity odd particles without undergoing any annihilation. The baryon and dark matter abundances are mainly controlled by the dilution factor from moduli decay, which is typically in the range $10^{-9}-10^{-7}$. The exact number densities are determined by simple branching fractions from modulus decay, which are expected to be of similar order in the absence of symmetries. This scenario can naturally lead to the observed baryon asymmetry and, for moderate suppression of the two-body decays of the modulus to $R$-parity odd particles, can also yield the correct dark matter abundance for a dark matter mass in the $(5-500)$ GeV range.

\end{abstract}
MIFTP-10-48
\maketitle


\section{Introduction}

Recently there has been significant interest in exploring non-standard thermal histories of the universe. In a standard thermal history, the universe reheats after inflation and then cools adiabatically during its subsequent evolution. In non-standard thermal histories, on the other hand, there is additional entropy release from the decay of some beyond the standard model particles (moduli fields from string theory being a famous example) that dominate the universe at some stage. These decays can happen at various time scales with consequences ranging from benign to disastrous.

If the modulus reheats the universe (far) above the electroweak scale, its effect is rather mild. Thermal freeze-out of dark matter (which happens below the electroweak scale) will be essentially unchanged in such a scenario, and standard scenarios of baryogenesis (including leptogenesis) may also be brought into play after modulus decay.
At the other extreme where the reheat temperature from modulus decay is below MeV, it will ruin the success of Big-Bang Nucleosynthesis (BBN)~\cite{BBN}. This is the infamous cosmological moduli problem, first pointed out for the Polonyi field in the earliest hidden sector models for supersymmetry breaking~\cite{Banks:1993en, Banks:1996ea, Banks:1995dt}.

Moduli decay with a reheat temperature in the $1 ~ {\rm MeV}-1 ~ {\rm GeV}$ range, corresponding to a mass range $20 ~ {\rm TeV} \lesssim m_\tau \lesssim 1000 ~ {\rm TeV}$. It has been argued that string compactifications will generally have moduli in this range, and certainly explicit examples in the literature do (for example the simplest - KKLT~\cite{Kachru:2003aw}).

This fact can have interesting consequences. The entropy produced by the decaying modulus dilutes any dark matter and baryon asymmetry that have been generated previously. Moreover, modulus decay typically leads to non-thermal production of dark matter, and may also result in a baryon asymmetry. Thus the dark matter and baryon densities receive contributions from both the pre- and post-decay epochs.

The dilution factor due to entropy release by modulus decay, given by $Y_\tau = (3 T_r / 4 m_\tau)$, then takes a value in the $10^{-9}-10^{-7}$ range. This implies that any previously generated baryon asymmetry and/or dark matter will be hugely suppressed after the modulus decay. On the other hand, one may consider the case where modulus decay is entirely responsible for creation of baryon asymmetry and dark matter. This will render any pre-decay production of these quantities, which are suppressed by a large factor $10^7-10^9$, irrelevant~\footnote{In fact, this is the only possibility in scenarios where the modulus drives inflation~\cite{ADS1}.}.

In this paper, we propose such a non-thermal origin for baryon asymmetry and dark matter by noting that the small value of $Y_\tau$ is suggestive in two ways. First, it is remarkably close to the observed baryon asymmetry $\eta_B \sim 10^{-10}$. The $1-3$ orders of magnitude difference can be readily accounted for as follows. Late-time baryogenesis occurs by the decay of the modulus to some species $N$, which has $B-$ and $CP$-violating decays to the observable sector. It is reasonable to expect, from simple counting of degrees of freedom, that $N$ will be produced at the $1\% \,- \, 10\%$ level from moduli decay. In addition, the asymmetry parameter from $N$ decay, which is a one-loop effect, can readily supply another factor of $10^{-1}$. Thus one sees such non-thermal scenarios can accommodate baryogenesis naturally.

Second, the small value of $Y_\tau$ is also suggestive in regards to the dark matter-baryon coincidence puzzle
%
$\Omega_{\rm DM} \sim 6 \Omega_{\rm B}$.
%
For a dark matter mass within the $(5-500)$ GeV range, the dark matter number density (normalized by the entropy density) must be in the $\sim 10^{-12}-10^{-10}$ range. Suppressing modulus decay to $R$-parity odd particles (which eventually decay to the dark matter) by a factor of $\gtrsim 10^{-3}$, together with $Y_\tau \sim 10^{-9}-10^{-7}$, can then yield the correct dark matter abundance without any need for annihilation. (In fact, as we will see, dark matter annihilation after modulus decay is inefficient for reasonable values of the annihilation cross section.) The interesting point is that three-body decays of the modulus, which are suppressed by a factor of $\sim 10^{-3}$ due to phase space factors, inevitably produce $R$-parity odd particles at the required level. Therefore, all we need is to suppress the two-body modulus decays down to (nearly) such a level.



There have been many attempts in the literature to address the coincidence puzzle of baryon and dark matter densities~\cite{Thomas:1995ze, Kaplan:1991ah, Kitano:2008tk, Davoudiasl:2010am, Buckley:2010ui, Hall:2010jx, Haba:2010bm, Kaplan:2009ag, McDonald:2010rn}. These proposals relate the number density of dark matter and baryons, and hence reduce the problem to a prediction of dark matter mass. In this non-thermal scenario, too, we connect the number density of baryons to that of the dark matter particle.

The important point about our approach is that the abundance of baryons and dark matter are entirely determined by the couplings between the modulus sector and the observable sector since there is no dark matter annihilation. Rendering annihilation irrelevant is an advantage if one agrees that the physics of annihilation is not related to that of baryogenesis.
The abundances are then mainly controlled by the dilution factor from modulus decay, with the exact numbers depending on the branching fractions to $N$ and dark matter. In the absence of any symmetries, these branching fractions are expected to be of similar orders. This can naturally lead to the correct baryon asymmetry, and also address the coincidence problem for $(5-500)$ GeV  dark matter mass.


To be concrete, we consider the case where $R-$parity is conserved, and the Lightest Supersymmetric Particle (LSP) is the dark matter. Our examples in the modulus sector will be in the context of effective supergravity descending from string theory. The decay conditions will depend on the nature of the decaying modulus and the geometry of the compactification.

The rest of this paper is organized as follows. In Section \ref{Baryo}, we discuss the cosmological history, and the naturalness of baryogenesis in non-thermal scenarios. In Section \ref{Coin}, we describe our approach to the coincidence puzzle. In Section \ref{Modulus}, we outline the conditions on the modulus sector that lead to the desired branching ratios, relegating a more detailed discussion to the Appendices. We close the paper with conclusions.

\section{Baryogenesis} \label{Baryo}

As described in the Introduction, we consider a scenario where both the dark matter abundance and the baryon asymmetry of the universe are created from the late-time decay of a modulus $\tau$. We first briefly discuss the cosmological history of the modulus $\tau$.

\subsection{Cosmological History}

Cosmological moduli can dominate the energy density of the universe if they are displaced from the minimum of their potential~\cite{Giudice:2000ex}.
%
%
%
The decay width of the modulus is
\be \label{decaywidth}
\Gamma_{\tau} = \frac{c}{2\pi} \frac{m_\tau^3}{\Lambda^2} ,
\ee
where $c \sim 0.1-1$ and $\Lambda$ is the effective suppression scale for the interaction of the modulus $\tau$. Typically $\Lambda \, \sim \, M_p$, where $M_{\rm P} = 2.4 \times 10^{18}$ GeV is the reduced Planck mass.
However, we will also consider scenarios where the interaction scale $\Lambda \gg M_{\rm P}$ due to geometric effects (alternatively, one can think of these scenarios as having $c \ll 1$).
The decay occurs when $H \simeq \Gamma_{\tau}$, with $H$ being the Hubble expansion rate.
The universe becomes radiation dominated immediately after the modulus decay and the reheat temperature $T_r$ is given by
%
\begin{eqnarray} \label{Tr}
T_r \simeq
(5 ~ {\rm MeV}) ~ c^{1/2} \left(\frac{10.75 }{ g_*}\right)^{1/4} \left(\frac{m_\tau}{100~{\rm TeV}}\right)^{3/2} \left(\frac{M_{\rm P}}{\Lambda}\right) \, , \nonumber \\
\end{eqnarray}
where $g_*$ is the total number of relativistic degrees of freedom at $T = T_r$ ($g_* = 10.75$ for $T_r \gtrsim {\cal O}({\rm MeV})$). The dilution factor for $\tau$ decay is then found to be
%
\bea \label{Yt1}
Y_\tau \sim
(5 \cdot 10^{-8}) ~ c^{1/2} ~ \left(\frac{m_\tau}{100\, {\rm TeV}}\right)^{1/2} ~ \left( \frac{M_{\rm P}}{\Lambda} \right) \, ,
\eea
Considering a modulus mass of $(100-1000)$ TeV, with $c \sim 0.1-1$ and $\Lambda = (1-10^3) M_{\rm P}$, one finds
\be \label{Yt2}
10^{-9} \lesssim Y_\tau \lesssim 10^{-7}.
\ee
%

\subsection{Naturalness of Baryogenesis}

Here we study the case where $\tau$ is a string modulus.
Then a generic scenario for baryogenesis is to have $\tau$ decay to a species $N$, which has $B$- and $CP$-violating, but $R$-parity conserving, couplings to the observable sector fields~\footnote{We note that there are other options as well, such as using Affleck-Dine baryogenesis~\cite{Dine:1995kz, Affleck:1984fy} to produce $\mathcal{O}(10^{-2})$ asymmetry, or having direct $B-$violating couplings of the modulus to the visible sector. In the first case, one has to worry about initial conditions~\cite{Dutta:2010sg}. In the second, conserving $R-$parity typically requires that the vev of the modulus vanishes, which is difficult for geometric moduli we will study. Moreover, the coincidence problem is less direct to address in those scenarios.}.

Denoting the branching ratio for $\tau$ decay to $N$ by ${\rm Br}_N$, the baryon asymmetry of the universe
%
%
is given by
\be \label{BAU}
\eta_B \equiv {n_B - n_{\bar B} \over s} =
Y_\tau  ~ {\rm Br}_N \, \epsilon \,\, ,
\ee
where $\epsilon$ is the generated asymmetry per $N$ decay.
%
As mentioned before, see Eq.~(\ref{Yt2}), we have $10^{-9} \lesssim Y_\tau \lesssim 10^{-7}$. The observed value of $\eta_B \simeq 9 \times 10^{-11}$ is then obtained for:
\be
\epsilon \sim 10^{-1} ~ ~ ~ , ~ ~ ~ 10^{-2} \lesssim {\rm Br}_N \lesssim 1.
\ee

For $\mathcal{O}(1)$ couplings and $CP-$violating phases between $N$ and matter fields,
one has $\epsilon \sim 10^{-1}$ due to the one-loop factor. The range $10^{-2}-1$ for ${\rm Br}_N$ is also typical for $N$ production form modulus decay.
For a democratic decay of the modulus to the observable sector, simple counting of the number of degrees of freedom suggests that ${\rm Br}_N \sim {\cal O}(10^{-2})$.
For colored particles the branching ratio can be larger $\sim 10^{-1}-1$.

A concrete model of baryogenesis along this line has been proposed in~\cite{Allahverdi:2010im}.
Here we briefly discuss one possibility for baryogenesis in this model. In addition to the the MSSM, two flavors of singlets $N_{\alpha}$ and a single flavor of colored triplets $X, \overline{X}$ (with hypercharges $+4/3,-4/3$ respectively) are introduced. The superpotential is given by
\bea \label{superpot2}
W_{\rm extra} = \lambda_{i\alpha} N_{\alpha} u^c_i X + \lambda^\prime_{ij} d^c_id^c_j \overline{X}
+ {M_{\alpha} \over 2} N_{\alpha} N_{\alpha} + M_{X} X \overline{X} \, . \nonumber \\
\eea
The interference between the tree-level and one-loop diagrams in the $N_\alpha \rightarrow X u^c$ decay generates a baryon asymmetry. For $\mathcal{O}(1)$ phases and couplings (which are allowed by experimental bounds), the asymmetry per decay is given by
\be
\epsilon \sim \frac{1}{8 \pi} \frac{[{\rm Tr} (\lambda \lambda^{\dagger})] ~ [{\rm Tr} (\lambda^{\prime} \lambda^{\prime \dagger})]}{{\rm Tr} (\lambda \lambda^{\dagger})} \sim 0.1 \,\,.
\ee
This, along with $Br_{N} \sim 10^{-2}$, yields the correct baryon asymmetry for $Y_\tau \sim 10^{-7}$.

On the other hand, one can have two flavors of triplets $X,\overline{X}$ and a single flavor of $N$. In this case $X \rightarrow N u^c$ decays create a baryon asymmetry.
Then ${\rm Br}_X \sim 10^{-1}-1$ gives rise to the correct baryon asymmetry for $Y_\tau \sim 10^{-9}$. We note that $10^{-9}$ is the absolute lower bound on $Y_\tau$ in order to obtain the correct baryon asymmetry without fine-tuning (such as invoking resonant baryogenesis).


\section{The Coincidence Problem} \label{Coin}

In this section, we discuss how our scenario can address the baryon-dark matter coincidence problem.

The number density of dark matter particles produced via thermal freeze-out prior to modulus decay is diluted by a huge factor $10^7-10^9$, and hence is negligible. The abundance of non-thermally produced dark matter right after modulus decay is given by
\be \label{DM}
{n_\chi \over s} = Y_\tau {\rm Br}_\chi ,
\ee
where ${\rm Br}_\chi$ denotes the branching ratio for decay of the modulus to $R$-parity odd particles~\cite{Moroi:1999zb}. Note that each of these particles will end up giving one dark matter particle via chain decay. The dark matter annihilation rate $\Gamma_{\rm ann} = n_\chi \langle \sigma_{\rm ann} v \rangle$ is less than the Hubble expansion rate $H(T_r)$ for $\langle \sigma_{\rm ann} v \rangle \lesssim 3000$ pb. Dark matter annihilation after modulus decay is therefore negligible for reasonable values of annihilation cross section~\footnote{This can also be seen from the fact that the final dark matter abundance is given by the minimum of ${\rm Br}_\chi Y_{\tau}$ and $(45/8\pi^2 g_*)^{1/2}(M_{\rm P} T_r \langle {\sigma_{\rm ann} v} \rangle )^{-1}$. This is equal to ${\rm Br}_\chi Y_\tau$ unless the annihilation cross section is exceptionally large (which has been utilized to explain the observed positron excess in the PAMELA data~\cite{Acharya:2010af}.}. Therefore the expression in Eq.~(\ref{DM}) represents the dark matter abundance in our scenario.

The baryon and dark matter density ratio then takes the form
\be
\frac{\Omega_{\rm B}}{\Omega_{\rm DM}} \, \simeq \,
\frac{1 ~ {\rm GeV}}{m_\chi} \times \frac{\epsilon \, {\rm Br}_N}{{\rm Br}_\chi} \,\,.
\ee
%
Without specific symmetries forbidding certain decay modes one expects the two branching ratios $Br_\chi$ and $Br_N$ to be of orders. Thus this scenario can address the coincidence problem for a dark matter mass around the weak scale.

The values of $\Omega_{\rm DM}$ and $\Omega_{\rm B}$ are mainly controlled by the dilution factor from moduli decay, with the actual values determined by the corresponding branching ratios.
%
%
%
%
%
%
%
%
For the typical range of $Y_\tau$ given in Eq.~(\ref{Yt2}), the desired dark matter abundance $\Omega_{\rm DM} \simeq 0.24$ is obtained for
\be
{\rm Br}_\chi \gtrsim 10^{-3} ~ ~ ~, ~ ~ ~ 5 ~ {\rm GeV} \lesssim m_\chi \lesssim 500 ~ {\rm GeV}.
\ee
This requires the two-body decays of the modulus to $R$-parity odd particles to be suppressed. Note, however, that these particles are inevitably produced from three-body decays of the modulus. For example, consider the decay mode with two gauge bosons in the final state $\tau \rightarrow g g$. This gives rise to three-body decay of the modulus to gauginos $\tau \rightarrow g {\tilde g} {\tilde g}$ when one of the gauge bosons is off-shell. It is striking that the three-body decays are suppressed by a factor of $\sim 10^{-3}$ relative to two-body decays based on the phase space factors.

Therefore the coincidence problem can be addressed in our scenario if two-body decays of the modulus to $R$-parity odd particles are suppressed down to (nearly) the level of three-body decays. In the Appendix, we give conditions on the moduli sector in order to have such a suppression.



\section{The Modulus Sector} \label{Modulus}

In this section, we summarize the geometric conditions for which $Y_\tau Br_\chi$ is small, relegating detailed discussions of particular scenarios of LSP suppression and specific decay modes of moduli to the Appendix. We discuss various identities of the last decaying modulus, in the context of $D=4,N=1$ supergravity descending from string theory.

$(1)$ K\"ahler modulus local to the visible sector: The typical value of $Y_\tau$ is $\sim 10^{-7}-10^{-8}$. It is possible to argue on general terms that two-body decays of the modulus to $R-$parity odd particles are subdominant except for the decay to gauginos and the gravitino. The decay to a gravitino is already suppressed at $\mathcal{O}(10^{-2})$ due to phase space. Moreover, it will typically suffer kinematic suppression, if the modulus is near the gravitino mass. The decay through the gaugino channel will be subdominant if certain conditions on the geometry are met~\footnote{A modulus $T$ couples to the gauginos (as well as Higgsinos) through a term proportional to $\partial F^T$, which should be $\mathcal{O}(10^{-1}m_T)$. Typically this would require a compactification with multiple K\"ahler moduli, and some conditions on the K\"ahler potential.}. Modulus
decay to the $R$-parity even components of $X, \bar X$ and $N$ are not suppressed which produces baryogenesis as discussed above.

$(2)$ A non-local geometric modulus: In a typical compactification, there will be K\"ahler moduli that are not local to the visible sector. These moduli decay into visible sector fields with a width that is further suppressed with respect to the Planck scale due to geometric effects. From a calculational point of view, this happens because mass eigenstates corresponding to non-local moduli suffer volume suppression in their canonical normalization~\cite{Conlon:2007gk}. For non-local moduli, the dilution factor can be as small as $Y_\tau \, \sim \, \mathcal{O}(10^{-9})$.

$(3)$ A hidden sector scalar modulus: If the final decaying modulus is a hidden sector scalar $\sigma$, the main decay channels are to visible sector sfermions and Higgs bosons. It is possible to construct cases where the modulus decays mainly to Higgs bosons~\cite{Acharya:2008bk} and in this case the LSP production is suppressed. On the other hand, if the local K\"ahler metric for the visible sector depends on $\sigma$, the relative branching to $R-$parity odd and even particles depends on derivatives of $F^{\sigma}$, which are suppressed if the modulus receives non-supersymmetric mass.

\section{Conclusion}

In this paper, we have examined the coincidence between the observed dark matter and baryon densities of the universe in the context of late-decaying moduli with a mass $\sim 100$ TeV.
The main point of this proposal is the following:
The dark matter and baryon asymmetry are both directly produced from a common source, i.e., the decay of a modulus. In particular, the dark matter annihilation is irrelevant for all reasonable values of annihilation cross section. The modulus yield $Y_\tau = (3 T_r/4 m_\tau)$ is the main factor controlling the baryon asymmetry and the dark matter abundance. It typically takes a value in the $10^{-9}-10^{-7}$ range. The exact number densities of baryons and dark matter are determined by simple branching fractions from modulus decay. They are expected to be of similar order in the absence of symmetries.

Specifically,  we have outlined a natural scenario where baryogenesis occurs by modulus decay to some species $N$, which has $B-$ and $CP$-violating decays to the observable sector. Within the expected range $10^{-2} \lesssim {\rm Br}_N \lesssim 1$, the correct value of baryon density is obtained for ${\cal O}(1)$ couplings and $CP-$violating phases between $N$ and the matter fields.

In this secnario, dark matter is produced via chain decay of $R$-parity odd particles that are created from modulus decay. The correct abundance is obtained within the mass range (5-500) GeV, when the corresponding branching ratio satisfies ${\rm Br}_\chi \gtrsim 10^{-3}$. The lower bound is saturated by three-body decays of the modulus, and hence all needed is to suppress the branching ratio for two-body decays of the modulus to $R$-parity odd particles down to this value. We discussed various ways in which such a suppression can happen for string moduli.



Thus the hierarchy between the moduli mass and the reheat temperature, which is due to the gravitational coupling of moduli to matter, can be turned into a virtue and address the two major problems in cosmology.

\section{Acknowledgement}

We would like to thank Scott Watson and KS would like to thank Tom Banks, Takeo Moroi and David Morrissey for useful discussions. This work is supported by the DOE grant DE-FG02-95ER40917.

\appendix
\section{Scenarios of LSP suppression in Modulus Decays}

We work in the context of $D=4,N=1$ effective supergravity.

\subsection{Decay of a Local Modulus}

The coupling of a local modulus to the visible sector is governed by several terms~\cite{Kaplunovsky:1993rd, Brignole:1993dj}:

$(i)$ The gauginos and gauge bosons couple through the gauge kinetic function. We consider a scenario in which the visible sector is constructed on D7 branes wrapping a cycle $\Sigma$, with gauge coupling given by $1/g^2 = V(\Sigma)$, where $V(\Sigma) = {\rm Re} T = \tau$ is the volume of $\Sigma$ in string units.

$(ii)$ Visible sector fermions and scalars couple to the modulus through the Kahler potential and soft terms.

$(iii)$ The gravitino couples to the modulus.

The decay modes of the modulus are discussed in Appendix B. Here, we summarize the relevant conditions.

Concretely, the parameters for the moduli is given by
\bea \label{KW}
K \, &=& \, -2\ln \mathcal{V} \nonumber \\
W \, &=& \, W_{\rm flux} + \sum_i^N A_i e^{-a_iT_i} \nonumber \\
f_i \, &=& \, \frac{T_i}{2\pi}  \,\,\,.
\eea
In the above, $\mathcal{V}$ is the volume of the Calabi-Yau manifold. The superpotential receives two contributions: $W_{\rm flux}$ which fixes complex structure moduli, and a non-perturbative contribution given by gaugino condensation on $D7$ branes.

The modulus coupling to matter fields in the Kahler potential is assumed to be of the following form
\be
K \, \supset \, \widetilde{K}(\tau) \bar{\phi} \phi \,\, + \,\, Z(\tau)\, H_u H_d \,\,.
\ee
In a generic compactification with multiple moduli, it is necessary to go to the mass eigenbasis for the moduli, as well as canonically normalize the kinetic term. The normalized eigenstates $\tau_n$ are given by
\be \label{norm}
(\tau)_i \,\, = \,\, \sum_j \,\, C_{ij} \,(\tau_n)_j \,\,\,,
\ee
where the $C_{ij}$ are eigenvectors of the matrix $K^{-1} \,\partial^2 V$.


We will require that the modulus decays mainly to gauge bosons, possibly at a rate comparable to the Higgs boson and the $R$-parity even components of $X, \bar X$ and $N$. The gaugino/Higgsino channel, on the other hand, is sub-dominant at a level $\mathcal{O}(10^{-2}-10^{-3})$. These channels turn out to have decay widths that are proportional to derivatives of $F^{T}$.

The conditions for acceptable reheat and suppression of decay to gauginos and gravitinos are obtained from Eqs.~(\ref{norm},\ref{gauginodecay},\ref{GammaTGravitino}), respectively, as ratios of the decay amplitudes of the various channels:
\bea \label{conditions}
\frac{1}{\Expect{\tau}} C_i  & \sim & 1 \nonumber \\
\sum_{p(\bar{p})} \frac{C_{p(\bar{p})} (\partial_{p(\bar{p})} F^{T_i})}{m_{T_i}} & \sim & 10^{-1}-10^{-2} \nonumber \\
\frac{m_{T_i}}{m_{3/2}}|G_{T_i}| K_{T_i\bar{T_i}}^{-1/2} & \sim & 0.1-0.5 \,. \nonumber \\
\,
\eea
Here, we have assumed that the $p^{th}$ modulus is predominantly aligned along a single normalized eigenstate $(\tau_{n})_p$, with a coefficient $C_p$. Note that there is no sum over $i$. Gaugino production from kinetic terms are suppressed. Note that gravitino production would also be kinematically suppressed if moduli have mass $\sim \mathcal{O}(m_{3/2})$.

From the data in Eq.(\ref{KW}), we can compute the derivatives of the F-terms
\bea
\partial_{p} F^{T_i} \, &\sim & \,- m_{3/2} \left(\sum_j\partial_p K^{i\bar{j}}\partial_{\bar{j}}K + 2\delta_{ip}  \right) \nonumber \\
\partial_{\bar{p}}F^{T_i} \, &\sim & \, - m_{3/2} \left(\sum_j\partial_p K^{i\bar{j}}\partial_{\bar{j}}K + 2\delta_{ip}  \right) \nonumber \\
&+& m_{3/2}\left( 2\tau_i + K^{i\bar{p}}a_p  \right) \partial_{\bar{p}}K \,\,.
\eea
In the above, we have defined $a_p = \frac{\partial_{\bar{p}} \partial_{\bar{p}} \bar{W}}{ \partial_{\bar{p}}\bar{W}}$. Note that repeated indices are not summed over unless explicitly mentioned.

In  the simplest case where the moduli are decoupled from each other in the K\"ahler potential and appear in power-law, one has $\partial_p K \sim (1/\tau_p)K$, $\partial_p K^{i\bar{j}} \sim (1/\tau_p)K^{i\bar{j}}$, and the dominant behavior in the above is roughly $\partial_p F^i \sim 2m_{3/2} \delta_{ip}$, $\partial_{\bar{p}}F^i \sim 2m_{3/2} \delta_{ip} + m_{3/2}(a_p K^{i\bar{p}})\partial_{\bar{p}}K$. Assuming $a \sim 1$, and $K^{i\bar{p}}\partial_{\bar{p}}K \gg 1$ which happens in the simplest case of KKLT, it is clear that the last term in $\partial_{\bar{p}}F^i$ dominates.

In that case, the supersymmetric mass of the $p^{th}$ modulus is given by
\be
m_p^2 \, \sim \, \sum_{i\bar{j}}K^{p\bar{p}}K_{i\bar{j}}\partial_{\bar{p}}F^i \partial_p F^{\bar{j}} \,\,.
\ee
Clearly, for a compactification with a single local modulus, one has $K^{p\bar{p}}K_{p\bar{p}} = 1$ and the gaugino production channel is unsuppressed, as has been noted by several authors in the context of KKLT~\cite{Endo:2006ix, Nakamura:2006uc, Dine:2006ii}. In the case of multiple moduli, the situation is more complex, and depending on the ratio $\sum_{i\bar{j}}K^{p\bar{p}}K_{i\bar{j}}$ as well as the normalization factors $C_p$, one may get suppression. Moreover, one can easily have $\partial F \, \sim \, F \, \sim \, \mathcal{O}(m_{3/2})$, if $a_p K^{i\bar{p}})\partial_{\bar{p}}K \sim 1$. In that case, the modulus mass receives other sizable contributions, from taking a derivative of the gravitino mass and a derivative of the K\"ahler metric. We leave a detailed study of these geometric conditions for future work.

\subsection{Decays of Non-Local Moduli}

An important feature that distinguishes non-local moduli from a local modulus is the fact that the overall decay coefficient $c$ for non-local moduli may be small due to geometric effects (in the previous subsection, we have considered $c \sim 1$ as is evident from Eq.(\ref{conditions})). Mathematically, this is equivalent to the statement that the effective interaction scale $\Lambda$ of the modulus with the visible sector is enhanced with respect to the Planck scale. While this leads to a lowering of the reheat temperature, we will consider cases where the mass of the modulus increases also, to keep the reheat temperature above BBN. The overall effect is a lowering of $Y_\tau$ at constant low-reheat $T_r$, as is evident below.

It is instructive to write the dependence of the dilution factor on the interaction scale $\Lambda$.
\be \label{YLambda}
Y_\tau \, \sim \, \left(\frac{T_{\rm r}}{5\, {\rm MeV}}\right)^{1/3} \, \left(\frac{M_p}{\Lambda}\right)^{2/3} \, \times 5c^{1/2}\, 10^{-8}
\ee
For $\Lambda \sim 10^3 M_{\rm P}$, one can have $Y_{\tau} \sim 10^{-9}$.

Physically, this is an effect of locality. A local modulus on which the visible sector is supported will promptly decay gravitationally into the visible fields. A non-local modulus which is separated from the visible sector in a large volume $\mathcal{V}$ compactification also faces suppression due to the bulk separation. The various local geometric moduli are decoupled from each other in the K\"ahler metric by powers of $1/\mathcal{V}$. In that case, going to the eigenbasis ${\phi_i}$ of $K^{-1}\partial^2V$, one obtains
\be
\tau_i \, = \, \mathcal{O}(\mathcal{V}^{p_i}) \, \phi_i \,\, + \,\, \sum_{j\neq i} \mathcal{O}(\mathcal{V}^{p_j}) \, \phi_j \,\,,
\ee
with $p_j < p_i$. In other words, the $i^{th}$ geometric modulus $\tau_i$ is aligned along the $i^{th}$ eigenvector $\phi_i$, and has components along other eigenvectors $\phi_j$ that are suppressed by powers of $1/\mathcal{V}$.

Considering the visible sector on $\tau_i$, we thus see that the reheat of $\phi_i$ is enhanced by a volume factor, and the relevant energy scale for decay is lower than the Planck scale (for example, it may be the string scale $\sim 10^{14}$). This decay has a high reheat temperature, and is relatively unimportant for our purposes. For the other eigenvectors, however, the reheat into the visible sector is suppressed by powers of $1/\mathcal{V}$. These decays occur at much lower temperatures, around $T \sim 1 \, {\rm MeV} - 1 \, {\rm GeV}$, although the BBN constraint can be avoided because the respective mass eigenstates are also enhanced by powers of $\mathcal{V}$. Moreover, these decays occur at an effective scale that is enhanced compared to the Planck scale by powers of $\mathcal{V}$. From Eq.~(\ref{YLambda}), one therefore obtains that the modulus abundance is lowered for such decays.

There are other requirements coming from ensuring

$(i)$ The reheat temperature does not fall below $1$ MeV, or go above $1$ GeV when thermal effects will become important.

$(ii)$ The abundance $Y_\tau$ does not fall below $\sim \mathcal{O}(10^{-9})$, to ensure natural baryogenesis.

$(iii)$ There is TeV-scale supersymmetry.

$(iv)$ There is compatibility with moduli stabilization. For example, stabilization may introduce hidden sectors and one must ensure that the modulus of interest decays primarily into the visible sector.

Clearly, for a fully realistic model one could ask for many other requirements. We will not present a detailed model, but rather mention the examples of $\mathbb{P}^4_{[1,1,1,6,9]}$ and $K3$-fibered 3-folds with volumes of the form
\bea
\mathcal{V} = \,\, \tau_1^{3/2} \, - \, \sum_i \, \tau_i^{3/2} \nonumber \\
\mathcal{V} = \,\, \sqrt{\tau_1}\tau_2 \, - \, \sum_i \, \tau_i^{3/2}
\eea
respectively, for which decay modes have been worked out explicitly recently~\cite{Cicoli:2010ha}. In both cases, a low reheat temperature may be obtained with $Y_\tau \sim 10^{-9}$ for certain choices of visible sectors. For example, in the $K3$-fibered case, one has $T_r \sim \mathcal{V}^{-10/3}$, $m_1 \sim \mathcal{V}^{-5/3}$, so that for $\mathcal{V} \sim 10^6$, one obtains $Y_\tau \sim 10^{-10}$ and $T_r \sim 10$ MeV.

\subsection{Hidden Sector Scalar Modulus}

It is interesting to consider the case where a hidden sector scalar $\sigma$ decays very late into visible sector fields. We will assume that the scalar has a non-zero F-term and contributes to supersymmetry breaking. The gaugino soft terms are typically dominated by the F-term of the local geometric modulus. In that case, these is negligible branching of $\sigma$ to gauginos~\footnote{ If the gaugino masses are dominantly due to $F^\sigma$, the modulus couples to them with widths proportional to $\partial F^\sigma$. If the modulus receives non-supersymmetric mass, then this channel is suppressed.}.

The couplings of $\sigma$ to other fields depends on to what extent the $\sigma$ sector is sequestered from the visible sector. In the limit of complete sequestering, the local K\"ahler metrics do not depend on $\sigma$ and thus there are no dimension five couplings through the K\"ahler potential. The decays then proceed through the sfermion and Higgs soft terms, and final branching ratios are somewhat model-dependent. It may be possible to preferentially decay to Higgs bosons - for example such a case has been obtained for a late-decaying hidden sector meson in $G2-$MSSM models~\cite{Acharya:2009zt}. We note that the condition depends heavily on the local K\"ahler data of the visible sector.

On the other hand, if the local K\"ahler metric has $\sigma$ dependence, the relative branching to $R-$parity odd and even particles depends on derivatives of $F^{\sigma}$, which are suppressed if $\sigma$ receives non-supersymmetric mass.


\section{Decay Modes of a Local Modulus}

\subsection{Decays to Gauge Bosons and Gauginos}

The dimension five operator governing the decay into gauge bosons is
\bea
\mathcal{L}_{\tau gg} &=& ({\rm Re} f) ~ \left( -\frac{1}{4}\mathcal{F}_{\mu\nu} \mathcal{F}^{\mu\nu }\right) \, \nonumber \\
&=& \frac{-1}{4M_{\rm P}} ~ \langle{\rm Re} f\rangle ~ \langle \sum_j \partial_{\tau_i} {\rm Re} f\rangle ~ C_{ij} (\tau_n)_j \mathcal{F}_{\mu\nu} \mathcal{F}^{\mu\nu } \, , \nonumber \\
\,
\eea
where we have suppressed gauge indices. For simplicity, we assume that $\tau_i$ is predominantly aligned along a single normalized eigenstate $\tau_n$, with a coefficient $C_i$. Thus, after canonically normalizing the gauge fields and the modulus, the decay rate for the process $\tau_i \rightarrow gg$ is
\be \label{GammaTgg}
\Gamma_{T_i \rightarrow {\rm gauge}} =  \frac{N_g}{128\pi} ~ \frac{1}{\Expect{\tau}^{2}} \, C_i^2 \frac{m_{T_i}^3}{M^2_{\rm P}}
\ee
where $N_g = 12$ is the number of gauge bosons. Note that for a single modulus model, the normalization factor $C_i^2 = K_{T\bar{T}}^{-1}$, so that $\frac{1}{\Expect{\tau}^{-2}} \, C_i^2 \sim 1$.

For gauginos, the relevant terms in the supergravity Lagrangian are
\bea \label{gaugino}
\mathcal{L}_{\tau_i \lambda \lambda} &=& {\rm Re}f \left( -\frac{1}{2} \bar{\lambda} \slashed{\mathcal{D}} \lambda  \right) \nonumber \\
&+& \frac{1}{4} ~ F^{i} ~ \partial_{i}f^{*} \bar{\lambda}_R\lambda_R + {\rm h.c.} \, \nonumber \\
\,
\eea
where $F^{i} \, = \, e^{K/2} ~ (D_{\bar{j}} W) ~ K^{i\bar{j}}$.

The decay rate through the kinetic term is suppressed by $(m_{\rm gaugino}/m_\tau)^2$. We note that the three-body decays to gauge boson and two gauginos is suppressed by $10^{-2} - 10^{-3}$.

However, the other piece contributing to two-body decay gives
\be
\mathcal{L}_{T_i \lambda \lambda} \supset \frac{1}{4M_{\rm P}} \sum_p \left(\Expect{\partial_{p} F^i} \, T_p + \Expect{\partial_{\bar{p}} F^{i}} \, \bar{T_p} \right) \bar{\lambda}_R\lambda_R + {\rm h.c.} \,\,.
\ee
As before, we assume that the normalized modulus is given by $T_p = C_p (T_n)_p$.

From this piece, the decay width of the modulus to gauginos is
\bea \label{gauginodecay}
\Gamma_{T_i \rightarrow \tilde{g}\tilde{g}} \, &=& \, \sum_p \frac{N_g}{128\pi} ~ C_p^2 ~ \Expect{\partial_p F^{i}}^2 \frac{m_{T_i}}{M^2_{\rm P}} \nonumber \\
\Gamma_{T^*_i \rightarrow \tilde{g}\tilde{g}} \, &=& \, \sum_p \frac{N_g}{128\pi} ~ C_p^2 ~ \Expect{\partial_{\bar{p}} F^{i}}^2 \frac{m_{T_i}}{M^2_{\rm P}} \,\, .
\eea
%


\subsection{Decay to Gravitino}

A geometric modulus decays to the gravitino through the following term
\begin{eqnarray}
\mathcal{L} &=& \frac{1}{4} \epsilon^{k\ell mn} \left( G_{,T_i} \partial_k T - G_{,T^*_i} \partial_k T^{*} \right)
\bar\psi_\ell \bar\sigma_m \psi_n \nonumber \\
&-& \frac{1}{2} e^{G/2} \left( G_{,T_i} T + G_{,T^*_i} T^{*}_i \right) \left[ \psi_m \sigma^{mn} \psi_n + \bar\psi_m \bar\sigma^{mn} \bar\psi_n \right], \nonumber\\
\end{eqnarray}
where $G =  K +\log |W|^2$ is the K\"ahler function. The decay width to helicity $\pm 1/2$ components is given by
\be \label{GammaTGravitino}
\Gamma_{T_i \rightarrow {\rm gravitino}} \sim \frac{1}{288\pi} ~ \left(|G_{T_i}|^2 K_{T_i\bar{T_i}}^{-1}\right) \frac{m_T^2}{m_{3/2}^2}\frac{m_{T_i}^3}{M^2_{\rm P}} \,\,.
\ee
%


\subsection{Decay to Visible Sector Fermions and Scalars}

The modulus decays to visible sector fields through couplings in the Kahler potential of the form
\be
K \, \supset \, \widetilde{K}(\tau) \bar{\phi} \phi
\ee
where $\phi$ is a visible sector chiral superfield and $\widetilde{K}$ is the Kahler metric for visible sector fields. After using equations of motion, these decays are proportional to
\be
\Expect{\widetilde{K}}^{-2} \Expect{\partial_{\tau}\widetilde{K}}^{2} \frac{m^2_{\rm soft}}{m^2_\tau} \, \sim \frac{m^2_{\rm soft}}{m^2_\tau}
\ee
and are suppressed.

There are also couplings from the soft terms in the Lagrangian, which are proportional to $(F^{T})^2 \, \sim \, m_{3/2}^2$ and are suppressed.

\subsection{Decay to Higgs and Color Triplets $X$}

There are two-body decays of $\tau$ to Higgs and other scalars through dimension five operators in the K\"ahler potential. Such interactions take the form
\be
K \,\, \supset \,\, \widetilde{K}(\tau) \bar{\phi} \phi \, + \, Z(\tau)\, H_u H_d \,\,.
\ee
In the above, $Z$ is a function of the local modulus for the cycle supporting the visible sector superfields. We have suppressed flavor indices. The decay width is given by
\be
\Gamma \,\,\sim \,\, \frac{1}{8\pi} ~ C_i^2 ~ \frac{1}{\widetilde{K}^2}\left(\partial_{i}\,Z\right)^2 ~  \frac{m_{T_i}^3}{M^2_{\rm P}} \,\,.
\ee
There is also branching to the fermionic superpartners. The decay width is given by
\be
\Gamma \,\,\sim \,\, \frac{1}{8\pi} ~ C_i^2 ~ \frac{1}{\widetilde{K}^2}\left(\partial_{i}\,Z\right)^2 ~ (\partial F^{T_i})^2 ~ \frac{m_{T_i}}{M^2_{\rm P}} \,\,.
\ee
This channel is somewhat similar to the gauge/gaugino channel, and the relative production of bosons and fermions depends on the derivative of the F-term. As a whole, the channel may be more or less important than the gauge/gaugino channel, depending on the local K\"ahler metrics $\widetilde{K}$ and $Z$.


\end{document}